\documentclass[11pt,twoside]{article}
\usepackage{asp2004}
\usepackage{psfig}
\usepackage{epsf}
\usepackage{graphics}
\usepackage{lscape}
\def\edcomment#1{\iffalse\marginpar{\raggedright\sl#1\/}\else\relax\fi}
\reversemarginpar

\begin{document}
\title{Spectroscopy and Near-Infrared Photometry of the Helium Nova V445 Puppis}
\author{Patrick A. Woudt$^1$ and Danny Steeghs$^2$}
\affil{$^1$ Department of Astronomy, University of Cape Town, Rondebosch 7700,
South Africa}
\affil{$^2$ Center for Astrophysics, MS-67, 60 Garden Street, Cambridge, MA 01238, USA}

\begin{abstract}
Nova Puppis 2000 (V445 Pup) has been proposed as the first example of a helium
nova. Recent optical spectroscopy of V445 Pup at V = 19.91 mag
obtained with IMACS on the 6.5-m Magellan telescope, shows that the 
spectrum consists of He\,{\small I}, [O\,{\small I}],
[O\,{\small II}] and [O\,{\small III}] emission lines and no hydrogen is present.
The spectroscopy shows an expanding nova shell with blue- and redshifted
velocity components around $\pm 850$ km s$^{-1}$ and $\pm 1600$ km s$^{-1}$. 
Images taken with Magellan under very 
good seeing conditions (FWHM $\sim 0.6''$) shows V445 Pup to be extended (full width at
zero intensity $\sim 1.9''$)
and elongated (position angle $\sim 150^{\circ}$). 
We have followed the
secular evolution of V445 Pup since the decline from (optical) maximum, 
at near-infrared  wavelengths ($J$, $H$ and $K_s$) using
the Infrared Survey Facility (IRSF) at the Sutherland site of the South African Astronomical
Observatory. We find that V445 Pup is still covered by a dense dust shell
more than three years after its outburst.
\end{abstract}


V445 Pup, the very first example of a helium nova (Ashok \& Banerjee 2003; Kato \& Hachisu 2003),
is unique in many respects; it is hydrogen-deficient (Wagner, Foltz \& Starrfield~2001; 
Ashok \& Banerjee 2003), enriched in helium and carbon, and the initially formed optically 
thin dust shell has developed into an optically thick dust shell (Henden, 
Wagner \& Starrfield 2001; Woudt 2002) which still obscured the nova in 2004 August.

Kato \& Hachisu (2003; these proceedings) modelled the optical light curve of V445 Pup
and deduced various (model-dependent) parameters for the system: a mass of the primary ($M_1$)
$\ge 1.33$ M$_{\odot}$, a mass-transfer rate ($\dot{M}$) of several times $10^{-7}$ M$_{\odot}$ 
yr$^{-1}$, and an estimated recurrence time of $\sim 70$ yr (based on the ignition mass of
several times $10^{-5}$ M$_{\odot}$). The short recurrence time between nova outbursts and
the weak photospheric wind, results in the white dwarf continuously growing in mass. 
This makes the high-mass white dwarf of V445 Pup a prime candidate progrenitor of a type Ia supernova. 
Alternatively, it could end up via accretion-induced collapse as a neutron star (Kato \& 
Hachisu 2003). Whatever the outcome in the evolution of V445 Pup, this nova provides
an important opportunity to study the secular behaviour of a helium nova.

Determining the distance to V445 Pup is of importance to distinguish between the three
likely possibilities for the parent population of these helium novae: 1.~ultracompact
binaries of the AM CVn-class of cataclysmic variables (CVs) (see Warner (1995) and Nelemans 
et al.~(2004)), 2.~CVs with helium donors (Iben \& Tutukov 1991), or, 3.~a high $\dot{M}$ 
hydrogen accretor.

\begin{figure}[!t]
\plotfiddle{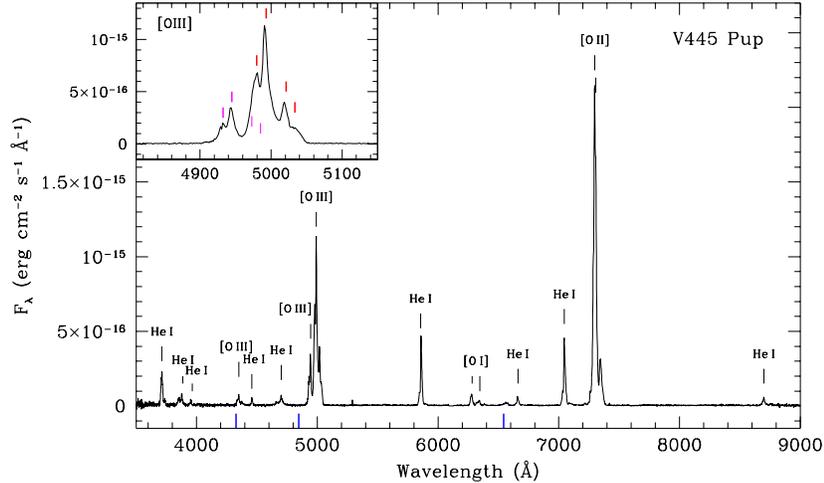}{6.0cm}{0}{55}{55}{-165}{-100}
\caption{An optical spectrum of V445 Pup, taken on 2003 December 17 (three years after the helium
nova outburst) with the 6.5-m Magellan telescope and the imaging spectrograph IMACS. The 
spectrum consists of emission lines of He\,{\small I}, [O\,{\small I}], [O\,{\small II}] 
and [O\,{\small III}], blueshifted either around $-850$ km s$^{-1}$, or around $-1600$ km s$^{-1}$. 
The [O\,{\small II}] and [O\,{\small III}] lines also have redshifted components at
$+850$ km s$^{-1}$ and $+1600$ km s$^{-1}$, as illustrated in the zoomed view of the 
[O\,{\small III}] lines. Note the very weak continuum. The small vertical lines below the continuum
indicate where hydrogen lines ought to have been if present.}
\label{woudtf1}
\end{figure}

Our detection of an expanding nova shell (see Figure~\ref{woudtf1}) offers the opportunity
to determine the expansion parallax of V445 Pup through high angular-resolution imaging
in combination with long-slit spectroscopy. No hydrogen features could be detected, suggesting
extreme He/H abundances. For example, we find $EW$(5876{\AA})/$EW(H\beta) > 50$.

The near-infrared light curve of V445 Pup (Woudt \& Steeghs, in prep.) shows that 3.5 years 
after the outburst V445 Pup remained covered by a dense dust shell (2004 May: $J$, $H$, 
$K_s$ = 18.5, 15.8 and 12.9 mag, respectively). A very large colour excess was 
still present at that time: $E(J-K_s) \approx 4.8 \pm 0.1$ mag (1999 February (2MASS):
$J$, $H$, $K_s$ = 12.3, 11.9 and 11.5 mag, respectively).

\end{document}